# Dominant Design Prediction with Phylogenetic Networks

Youwei He[a,b,*], Jeong-Dong Lee[a,b], Dawoon Jeong[a], Sungjun Choi[a], and Jiyong Kim[a]

[a]Technology Management, Economics and Policy Program, Seoul National University, 1 Gwanak-ro, Gwanak-gu, 08826, Seoul, South Korea; [b]Integrated Major in Smart City Global Convergence, Seoul National University, 1 Gwanak-ro, Gwanak-gu, Seoul, 08826, South Korea

*Corresponding author's e-mail address: heyouwei2020@snu.ac.kr*

**Abstract**

This study proposes an effective method to predict technology development from an evolutionary perspective. Product evolution is the result of technological evolution and market selection. A phylogenetic network is the main method to study product evolution. The formation of the dominant design determines the trajectory of technology development. How to predict future dominant design has become a key issue in technology forecasting and new product development. We define the dominant product and use machine learning methods, combined with product evolutionary theory, to construct a Fully Connected Phylogenetic Network dataset to effectively predict the future dominant design.

**1. Introduction**

In the technological development process, products are carriers of technology, and their success is determined by market demand. New products, as the result of technological progress, bring new market demands. For example, Apple's smartphones and Tesla's electric cars have opened the market for smartphones and electric cars, respectively. Therefore, technological evolution has its own trajectory, followed by market selection. For many years, predicting the trajectory of technological evolution has always been the dream of the communities of engineering forecasting for new product development and innovation management. Due to technological development uncertainties, technological forecasting has always been an unattainable task. Correctly predicted technical breakthroughs have an impact not only on company growth, industry dynamics, and countries' economies but also on humanity's technological level. Inventive activity is inherently uncertain due to the multitude of factors and the complexity of their interactions. Reducing this uncertainty to predict the technological profile of successful products would undoubtedly assist businesses in making informed investment decisions. Proponents of evolutionary theory, which emphasizes the role of technological trajectory as a crucial determinant of innovation dynamics (Dosi, 1982; 1988), incorporate the notion that technological choices are dependent on non-technical factors such as institutional or economic elements. There are three instances where technological paths are shaped by non-technical reasons. First, although technological trajectories are path-dependent, they may be influenced by non-technological factors, as illustrated by the development of the QWERTY keyboard (David, 1985). Second, market selection might not be purely driven by technological performance; it could result from business success due to strategic decisions, despite lower technological performance (for example, JVC's VHS triumph over Sony's Betacam (Cusumano et al., 1992). More broadly, it is widely recognized that the invention process is affected by cognitive and social phenomena, which help to clarify why the search for technology is an uncertain endeavor (Fleming, 2001). This study, by focusing solely on the technological profile of existing products to predict the success of future technologies, not only suggests a belief in technological determinism but, more importantly, we believe that the technological paths formed due to non-technical reasons are already reflected in our product data. Our machine learning approach, based on historical data, ensures that the influence of historical non-technical factors on prediction outcomes is also included within the model.

For a long time, to analyze the development of species over time, a phylogenetic network has been widely used in understanding the process of gene transfer between species in biology. In the process of product and technology evolution, we can also use this species evolution concept for this study. The framework of the phylogenetic network is based on panel data and clusters individual species in the same period through similarity to form a population. As time passes, the relationship between the previous and next generations is determined based on similarity. In a conventional phylogenetic network, each node represents a taxon. The directed connection between the nodes represents the evolutionary relationship between the ancestor and descendent generations. Although each node has only one ancestor, it can have many descendants. Technology and biology evolutions have many common aspects. Phylogenetic networks can also be used to describe how technologies evolve. A technical product phylogenetic network, which is proposed by Lee et al. (2022), is a phylogenetic network generated by using technical information contained in products. It can describe product evolution.

Nevertheless, the analytical framework of phylogenetic networks in biology cannot be directly applied in the context of technological development. For example, a conventional phylogenetic network in biology cannot address the situation of multiple ancestors in technology evolution as they are often in the form of trees. Moreover, some important genes of technology are not necessarily inherited from the ancestor. However, they are likely to be inherited from other predecessors with little similarity. Because a conventional phylogenetic network only retains the most similar connections, we cannot see the links with other predecessors in this network. In addition, for the prediction of technological trajectories, the population in this research need not be clustered as clustering focuses on common information and neglects different types of minor information. However, minor information can be crucial in predicting the dominant product in this study. Therefore, based on a conventional phylogenetic network, we build a fully connected phylogenetic network (FCPN) in this study. Although previous research adopted a directed acyclic graph (DAG) to address the problem of multiple ancestors of phylogenetic networks, the DAG is only a subset of a fully connected framework. The main function of a conventional phylogenetic network is population development visualization. However, the data of a fully connected phylogenetic framework are constructed for technical prediction in this study. Using a fully connected phylogenetic graph neural network (FCP-GNN), we build a model that can predict the future dominant product. The model predicts whether a product can incorporate technologies that are likely to become dominant in the future. That is, we aim to predict whether brand-new technology in a product can become dominant in the future at the time of the technology's birth. This has an evident guiding significance for the scientific and technological strategic decision-making of enterprises or countries.

In a conventional method, the correlation of low-dimensional features is first calculated, a certain threshold is set, and relevant features that cannot reach the threshold are eliminated. In this case, some features that play a key role in high-dimensional features are also eliminated. However, in this study, all low-dimensional correlations are retained, and high-dimensional correlations accordingly are mined. In addition, the phylogenetic network is used as a graph to be the input, rather than simple structural panel data or time-series data. Because input data are based on a graph, it is an organic whole, rather than scattered points. The input graph must have a form suitable for the prediction of the subsequent step.

The remainder of this paper is structured as follows. Section 2 briefly covers some relevant works and proposes the research question. Section 3 introduces the framework of technological evolutionary phylogenetic networks and our research objective. Section 4 describes the general research process and data and introduces the method. Section 5 performs a case study of the dominant design prediction of Chinese mobile phones and then examines the model's validity. Finally, Section 6 summarizes our important results and future research.

## 2. Literature and research question

### 2.1. Literature

Technological evolution is a theory that explains the mechanisms for the process of technological change. It follows biological evolution, which has its root in the book "On the Origin of Species," written by Charles Darwin. In biological evolution, the information being replicated and inherited from an ancestor to the next generation is defined as genotypes. In Farrell's (1993) study, considering the analogy between biological evolution and similar processes in technological evolution, the heuristic principle of "Generalized Darwinism" can explain aspects of technological development. In the product space, all the products comprise technologies as their attributes. By improving their attributes, they can have their evolutionary process and can form a product network based on their similarities among generations. Further, products have an evolutionary process among generations. Their evolutions are formed by changing the combination of different technologies. By the interaction of technologies transferring between products, an evolutionary product network can be formed. Every product can be regarded as a node in this network, and this network can expand over time. Hence, this process is the product of technological evolution. In this paper, we define product genotypes as the technologies, which are included in the product. The genotype can represent the technical characteristics of the product. In biology, a chromosome is a long DNA molecule with part or all the genetic material of an organism. We define the product chromosome as a string that can represent all the technologies in the products.

According to Utterback and Abernathy (1975), a dominant design is the one that is favored by the market, the one to which rivals and innovators must conform if they want to command a considerable market share. While Klepper (1996, 1997) suggests that the shift from an exploration phase to a shakeout is dictated by economic factors associated with the industry's demand structure, the concept of a dominant design resolves technological uncertainty. This resolution fundamentally changes the nature of innovation, transitioning from product to process innovation, and significantly impacts the competitive dynamics of the industry. A dominant design is a new product that is formed and integrated with multiple technologies. The emergence of the dominant design establishes a technological track for a certain product category; thus, other technological tracks are rejected by the market. Anderson and Tushman (1990) empirically suggested the operation definition of a dominant design; that is, a single or a narrow range of configurations that accounted for over 50% of new products or process installations and maintained a 50% market share for more than 4 years. According to Suarez et al. (2015), the dominant category is that most stakeholders adhere to the conceptual schema when referring to products that address similar requirements and compete for the same market space. From Abernathy and Utterback's (1978) opinion, dominant design is a turning point to the emergence of industrial changes by standardization and enables manufacturers to set up a mass-manufacturing system, which is also issued by Murmann and Frenken (2006). By using data on mobile phones from 1990 to 2003, Koski et al.(2007) analyzed the emergence of a dominant design and found that, dominant design forms only in horizontal but not vertical product features. In this paper, we defined the dominant technology (genotype) as the technology, which was contained by most products on the market and detect it by using a threshold of 50% of the new products which contain this technology in one specific year. This only followed Anderson and Tushman's (1990) operational definition. We also defined a new operational concept of the dominant product. When certain technology first appears on the market, it is contained in product A. If this technology becomes a dominant technology in the future, we defined product A as a dominant product. There is no limit to how many years later Product A can become dominant. However, from the result of this study, on average, it is around 3 years in the Chinese mobile phone dataset. We make the definition of similar dominant products because we aim to detect the dominant technology by the year of its birth, rather than the year when it becomes dominant, i.e., over a specific market share (i.e., 50%). We designed the concept of a "dominant product" only for the technology prediction. Its actual meaning is a product that contains a future dominant technology. Therefore, our "dominant product" concept is different from a conventional concept of a dominant product, which refers to a product that owns over 50% market share. If we predict a product as "dominant," this implies that this product has some specific technologies that will become dominant in the future. Specifically, predicting our "dominant product" is more meaningful than a conventional dominant product as the conventional dominant product is a fact and does not need to be predicted.

## 2.2. Machine learning and prediction research in technology evolution and research question

In recent years, machine learning has been greatly developed as a data analysis method. Although machine learning methods have been used widely in many fields, the progress in the field of technology evolution prediction is relatively slow. Wang et al.

(1999) were the first to propose the possibility of predicting future technology with artificial neural networks (ANNs). In their study, they used a combination of chaos theory and ANN to perform forecasting. Unfortunately, they provided only a conceptual framework without empirical verification. As Markou and Singh (2003) mentioned, although a few novelty detection models used a neural network, they were not related to technological evolution and dominant design prediction. Our model attempts to adopt machine learning to predict future technology based on phylogenetic networks. Many machine learning methods can capture non-linear relationships in a complex system, such as support vector regression (SVR) and decision trees, which use statistical methods to handle classification problems with some explanatory power. However, conventional machine learning methods' accuracy cannot compare with that of deep neural networks (DNNs) with big data. Biswas et al. (2022) forecasted the specification of the most influential features of future products by ANNs. However, this approach can only forecast the technological improvement in the same technological features and cannot handle new features that come out every year. Our study employed a DNN-based model approach to improve prediction capabilities and it can predict dominant products using not only the existing technological features but also information about new technologies that come out every year.

Instead of the conventional machine learning methods for prediction of the technology evolution, we used the technological phylogenetic network combined with graph neural network (GNN) models to obtain a better predictive capability. Our research question is how to construct the technological phylogenetic networks and combine them with the GNN to predict future dominant products.

## 3. Conceptual framework and research objective

*3.1. Technological evolutionary phylogenetic networks*

A product's evolutionary process is driven by a combination of the genotypes transferred from the previous generation and the newly created genotypes. New products are formed by these two types of genotypes. The genotypes transferring from the previous generation include vertical transmission (VT) and horizontal gene transfer (HGT), which have been discussed in Gyles and Boerlin's (2014) study. These newly created genotypes refer to mutation. The dynamic movement and combination of all these genotypes create new products, and the chronological evolution of all products can be represented as a phylogenetic network. Specifically, the technological product evolutionary phylogenetic network is the foundation of our study.

A phylogenetic network is a graph that can visualize the evolutionary relationships between nodes. This concept follows a rooted phylogenetic tree, which can present the evolutionary history of a set of species in the biological field. In a biological phylogenetic network, nodes represent species, but the chronological order is inaccurate. Meanwhile, the links between species are kinship. A technological phylogenetic network has exact chronological order, and the nodes can be a product or a taxon, which is a cluster of products. The links in the technological phylogenetic network are calculated based on similarities between the nodes in adjacent generations. The phylogenetic tree is a type of phylogenetic network. According to Huson and Bryant (2006), the difference between them is that, in a phylogenetic tree, every node can only have a single ancestor, whereas, in the phylogenetic network, nodes can have multiple ancestors. Huson and Scornavacca (2011) describe a graph using the concept of the DAG, indicating a directed graph devoid of cycles. In this paper, we defined the FCPN. In the FCPN, 1 year is regarded as a generation. All the nodes in each generation have fully directed connections to each node in the next adjacent generation. However, they do not have connections with nodes in other generations, including the current generation. Technically, this network is fully connected within adjacent generations only, not with all the generations. Although no direct connections exist between non-adjacent generations, the nodes in non-adjacent generations can be connected by indirect connections. In technological evolution, the main point is that information or genotypes can be spread from generation to generation. In the FCPN, any genotype in any node can be transmitted to any node in subsequent generations. The structural difference between the FCPN and the phylogenetic tree is that the FCPN keeps all links among adjacent generations. Conversely, a conventional evolutionary phylogenetic tree keeps just the most possible connection between generations.

When building a conventional technological evolutionary phylogenetic tree, as proposed by Lee et al. (2022) to investigate the technology evolutionary process, two types of conventional evolutionary phylogenetic trees can be used: a product phylogenetic tree (Fig. 1) and a taxonomic phylogenetic tree (Fig. 2). [Fig. 1 near here] In the product phylogenetic tree (Fig. 1), every node is a product. The directed links between the nodes in adjacent generations show their ancestor-descendant relationship. The link is determined by calculating the similarity between the node and all the nodes in the previous generation. In the previous generation, only the node with high similarity has a link with the node. Therefore, the link is directed from the ancestor to the descendant. In the technology evolution, we defined the taxon as a species, indicating a group of products with similar technologies. Researchers must cluster similar products in every generation to form the taxon and find their ancestor taxon based on high similarity with the previous generation. In the technology evolution, a product can be composed of a set of technologies (genotypes), and a taxon is composed of a set of products that contain similar technologies. Thus, researchers can find products' families and lineage. Whether it is a conventional product phylogenetic tree or a taxonomic phylogenetic tree, each node can only have one ancestor. The pseudo-code for building the conventional phylogenetic trees is represented by Algorithm 1. [Algorithm 1 near here]

Unlike conventional phylogenetic trees, the FCPN does not specify one ancestor. Its structure can guarantee that any genotype of a previous generation can be spread to the next generation without loss and clustering nodes to form taxon is not necessary. Moreover, in conventional phylogenetic trees, as each descendant only has one ancestor by high similarity, the FCPN forms the link bottlenecks for gene transferring between adjacent generations, leading to less diversity. In the FCPN, no direct genotype transferring exists with non-adjacent generations. However, genotypes can use more indirect paths transferring to non-adjacent descendants. Fig. 2 shows an example comparison of a conventional phylogenetic tree and an FCPN. Evidently, the conventional phylogenetic tree and an FCPN have the same nodes, and the difference between them is the links. Because of these different links, their structures also vary. As also shown in the figure, a conventional phylogenetic tree is a subgraph of the FCPN. If we transform the FCPN to a conventional phylogenetic tree, we lose some links, leading to the loss of information. [Fig. 2 near here]

*3.2.Research objective*

This study aims to predict dominant products by using a deep learning model based on the FCP-GNN framework through product evolution structural data. Once a dominant design emerges, the functional model remains unchanged over an extended period, even though the components continue to evolve (Robinson et al., 2024). Therefore, predicting the dominant design holds practical significance for product design. One dominant product contains at least one future dominant technology. Our model can predict the dominant products that contain dominant technology in the birth year of this technology, rather than judging it as a dominant product after the technology it contains becomes dominant. For example, technology "a" is born in product "b" in the year 2010, and "a" becomes dominant in the year 2014. Further, there is product "c" in 2014 that contains "a." Specifically, "a" is a dominant genotype; therefore, "b" is a dominant product. However, "c" may not be a dominant product. To achieve this, our task is to train an FCP-GNN-based DNN model and then find the optimal parameters of the GNN in Eq. 2 and other parameters in the following DNN models for an accurate prediction. The GNN and DNN make up the prediction function. When we optimize the prediction function, we must minimize the sum of prediction losses. We can obtain the best parameters through optimization. As this is a classification task, the loss function is the cross-entropy function. After training the model, the prediction function enables us to predict the future dominant design.

## 4. Methodology

### *4.1. Assumptions*

As it incurs time for the technology to spread, a brand-new technology incurs time from being introduced to becoming dominant. To forecast whether a certain technology can become dominant, we must allow it enough time to grow. Then, we can verify if it can be dominant. If the observation period is relatively short, the technology that can become dominant in the future may be wrongly determined as non-dominant. In this case, we assume that the average growth time for a brand-new technology to become dominant is approximately three years on average. The reason we selected three years is explained in the "Dominant technologies" section. The detailed results of the dominant technology growing years are presented in the "Results" section. Therefore, we use the data from the last three years as our test dataset. Our splitting approach causes the training data and test data to be not independent and identical distribution (i.i.d.), which influences the result's accuracy. We assume that the data of the training data set and test data set are close to the i.i.d.

Furthermore, in the construction of the FCPN, we assume the technologies of the product, i.e. product attributes, are the genotypes, and one-hot encoding can be used to convert them into chromosomes for every product. Additionally, the chromosome which consists of a set of genotypes can represent mobile phone products. We also assume that all the products in the same year cannot transfer genotypes with one another. Moreover, we assume that genotypes can be transferred only through adjacent years, which means this transferring process can happen only from a generation in the year T to the generation in the year T+1 but not T+i (i=2,3,…,n). Considering every product as a node, we assume every node has at least one ancestor except the first generation. Furthermore, we can construct links (edges) between ancestors and descendants. Moreover, we assume the link's direction is from the ancestor to the descendant, and the link's property (weight) indicates their similarity.

*4.2.Data and Prepossessing*

In this study, we used Chinese mobile phone data to construct the FCPN as mobile phone information can be easily obtained from the Internet. Further, we selected the Chinese market as it is the largest single-country market for mobile phones. China is the biggest producer of mobile phones. Therefore, the time lag between production and sale is short, and there are several brands of mobile phones. We collected the information from one of the largest mobile phone information websites in China (mobile.zol.com.cn) by Web Crawler. We collected 3642 mobile phone products with attributes based on a 3-level hierarchical structure (Fig. 3). The first level shows the main parts of the mobiles, and the second level shows the sub-parts of the first level. Furthermore, 19378 attributes are present at the third level. We only used the third-level attributes. After cleaning the data, we obtained 3430 products with 11126 attributes (genotypes). The mobile phone dataset includes 21 years (from 2001 to 2021) of product data, but in 2021, the data were until the 15th of March, implying that the data of 2021 has information of only 2.5 months. The cleaned dataset was provided by He (2022). We used Python, TensorFlow, and Spektral to develop our algorithm. Spektral is a specialized package developed by Grattarola and Alippi (2020) for graph learning based on the TensorFlow framework. We split the dataset into a training dataset, a validation dataset, and a test dataset. We used the validation dataset to avoid overfitting. [Fig. 3 near here]

Among the 11126 genotypes, we detected the dominant ones. A dominant genotype is in one specific year, and its count is more than 50% of the total product count. We found out that there are 161 genotypes that are dominant (without repeating) in the 21 years considered. Owing to some reasons related to calculations, we define the "dominant product." The dominant product is a product that is the first to have at least one future dominant genotype. We found 657 dominant products in these 21 years. Then we classified the products into the dominant and non-dominant ones, which are our labels in machine learning. The detailed data are presented in Table 3.

First, we need to transfer these genotypes into binary numbers by one-hot encoding so that they can represent the technologies in the products. We can assign each product a fixed length one-row vector that only consists of 0 or 1 in each dimension, where different columns represent different genotypes in the vector. We use this vector as the product's chromosome in this research. Second, completely link to the products from adjacent years. We use the chromosomes of the products to calculate their similarity and set the similarity as the link weight. The entire connected network is our FCPN. This is a directed graph. We compute the Jaccard similarity of chromosomes in Eq. 1 between paired products to obtain link weights.

$$Similarity(A, B) = \frac{|A \cap B|}{|A \cup B|}, \text{ Eq. 1}$$

where A and B are chromosomes of various products. The entire process of building an FCPN is a transformation from unstructured data to graph form. Pseudocode is in Algorithm 2. [Algorithm 2 near here]

In this case, our FCPN is a homogeneous graph. All the nodes are products, and the genotype information is extracted from their attributes. Our objective is to predict the dominant design. To prepare the data to train the model, we detect the dominant genotypes first. The dominant genotypes refer to the genotypes that can be dominant in the future. This genotype is labeled as dominant only in the first year of its appearance but not in the following years. Subsequently, we assign the labels to dominant products. Dominant products are those that include the dominant genotypes. One dominant product must include at least one dominant genotype. Therefore, there may be more than one dominant genotype in one dominant product. Some dominant products may include the same dominant genotypes in an identical year. Thus, the dominant products include the genotype that can be dominant in the future, but we only label them in the year when the genotype appears. This is to avoid the situation where the machine learning algorithm can judge whether the genotype is dominant or not according to the number of annual growths of the genotype, as there is no information regarding the history of the genotype before its appearance. In other words, if we predict the dominant genotype in its dominant year, instead of its birth year, then the algorithm can predict the genotype using its historical information by simply detecting the annual growth trend of genotype count and product count ratio and then determining the dominant product. Therefore, our approach has more predictive power. A label is a binary number. If the product is dominant, the label is 1, and it is 0 otherwise. Fig. 4 provides an illustration of the method of identification of the dominant product. In year T, there is a product 1 in which the 3rd genotype value is 1. Furthermore, it is the first time that the 3rd genotype appears in the entire product set. In year T+1, the 3rd genotype's market ratio is 1/3 which is less than 50%. In year T+2, there are a total of five products. Among these, three of them have the 3rd genotype value 1. The ratio is 3/5, which is more than 50%. Therefore, in year T+2, the 3rd genotype becomes a dominant genotype, and hence, product 1 is a dominant product. [Fig. 4 near here]

As we already have the graph with node and link attributes, we can use the same to feed into the GNN model. But the nodes' attributes correspond to the chromosomes. As the count of genotypes is 11126, the length of the chromosome is 11126. The matrix consisting of products as rows and genotypes as columns is sparse. As the graph already has information on similarities in its structure, we do not need too much information about the chromosomes. Therefore, we use principal component analysis (PCA) to reduce the dimension of the chromosomes. We drop 70% of chromosome information (the amount of variance that can be explained) and change the chromosome into a 107-dimension one-row vector for each node. This approach is efficient, but in the extreme case, it only retains 2% of the original information, which has only two dimensions. In this case, more epochs are required to obtain the convergence for the GNN. The proportion of the original information reserved can be set by researchers arbitrarily. We choose to retain 30% of the chromosome information because it is more efficient for training. The pseudo-code for labeling the dominant genotypes is in Algorithm 3. [Algorithm 3 near here]

We split the dataset into three parts—training, validation, and test data. Specifically, training data include the products from 2001 to 2015. The validation dataset contains the products from 2015 to 2018. The test dataset has the products from 2019 to 2021. Then, for robustness, we split training, validation, and test datasets with different year periods in several ways and apply our method. As we used the single data model to perform the node classification task in GNN, we must use the whole graph as the input and therefore use masks for differentiating the different datasets. [Table 1 near here]

*4.3.Model description*

Scarselli et al. (2008) argued that the GNN is a supervised learning method based on a neural network for data processing by inputting graph data structures as features. As a product's technological information can be represented with the FCPN, which is a graph in the non-Euclidean domains. The graph is suitable to be handled with the GNN. The notations used in the analysis are summarized in Table 1.

When we apply GNN to the FCPN, the main idea is to extract the localized spatial features and build highly expressive representations from the graph. To achieve this, GNN uses the convolution operator in the spectral domain to filter the noise and extract the features. For the spectral approaches, it uses the principle of signal filtering. That is, the signal can be decomposed into different frequencies, filtered through different filters to remove the noise of certain frequencies, and then reverse synthesize a new signal. The product evolution can use this filtering mechanism, as in each generation, every node is composed of many genotypes, which also can be regarded as signals that are transferred from ancestor products to descendent products. These genotypes transfer between generations is applied through HGF or VT process, and the new genotypes enter the scene through the mutation process. Some genotypes can become dominant, whereas others cannot. The product that contains the future dominant technology is the dominant product in our definition. The GNN filters the case of forming the non-dominant products by learning the genotype transferring and generation process.

In this study, we use ChebNet, which is proposed by Defferrard et al. (2016), as the main method of the GNN. The graph convolution operation can be represented as

$$Cheb(x) = c_f \star x \approx \sum_{k=0}^{K} A_k T_k(\tilde{L})x, \quad \text{Eq. 2}$$

where $\tilde{L} = \frac{2}{\lambda_{max}} L - I_N$, $\lambda_{max}$ denotes the largest eigenvalue of L, the range of the eigenvalues in $\tilde{L}$ is [-1,1]. $A \in R^K$ is a vector of Chebyshev coefficients. $R^K$ refers to K-dimensional Euclidean space. The Chebyshev polynomials are defined as $T_k(x) = 2xT_{k-1}(x) - T_{k-2}(x)$, with $T_0(x) = 1 \text{ and } T_1(x) = x$. In this research, for simplification, we just consider the case when k=2. Therefore, the learnable diagonal matrix filters are the weights that refer to $c_f$. Our objective is to learn filter $c_f$ which can filter the noise genotypes from the chromosomes $x$ and pass the dominant genotype information to the next generation in the technological diffusion process. This process is not only a feature transformation process but also an embedding process for the nodes' features from graph to vectors by performing the convolution operation on them.

*4.4. Model structure*

The single GNN model we use is a static graph, but we prefer dynamic graphs because the genotype transferring, and technology evolution processes are dynamic. We construct a dynamic graph model by stitching three static graphs together as a time series graph and then feed this dynamic graph into a bidirectional LSTM layer, which was proposed by Graves and Schmidhuber (2005). We use the bidirectional LSTM layer because the technology is sometimes developed in a reverse causality relationship. For example, some technology is developed and embedded in a current product, but this technology is developed for another future product. In other words, the future product is the reason the current technology has been developed. Meanwhile, because of the current technology, future products can be developed. The dynamic graph constructive process is iterative. The first static graph is the original one in Eq. 5, and the next one is based on the previous graph with a one-time convolution operation. We apply a graph convolution operation to the current graph to obtain the next graph, implying that the graph, after one time message passing of each node from ancestors to descendants and updating their chromosomes, changes into the graph of the next generation. As the test data only have three years, we only construct the dynamic graph with three generations. Based on Eq. 2, we can use the Chebyshev convolutional layer function in Eq. 6 and Eq. 7 to generate new generations of the graph. Eq. 3 shows that a chromosome of the product is composed of $n$ genotypes. Eq. 4 deals with the process of how we extract feature information from the original one-hot encoding chromosome by PCA.

$$x_i = (g_{0,i} \| g_{1,i} \| g_{2,i} \| \cdots \| g_{n-1,i}), \quad \text{Eq. 3}$$

$$\tilde{x}_i = PCA(x_i), \quad \text{Eq. 4}$$

$$\tilde{x}_i^{(1)} = \tilde{x}_i, \quad \text{Eq. 5}$$

$$\tilde{x}_i^{(2)} = Cheb_1(\tilde{x}_i^{(1)}), \quad \text{Eq. 6}$$

$$\tilde{x}_i^{(3)} = Cheb_2(\tilde{x}_i^{(2)}), \quad \text{Eq. 7}$$

$$\tilde{x}_i^{(merge)} = \left(\tilde{x}_i^{(1)} \| \tilde{x}_i^{(2)} \| \tilde{x}_i^{(3)}\right), \quad \text{Eq. 8}$$

$$d_i = Gap\left(BiLSTM\left(\tilde{x}_i^{(merge)}\right)\right) = f_{PFCN}(x_i, \theta), \quad \text{Eq. 9}$$

where $i$ refers to product's index, and $\theta$ refers to the learnable parameters in the model.

Thus, in our model, we generate time series graphs for 3 generations. The first-generation graph is the original graph in which a product is represented by its chromosome in Eq. 5. The second-generation graph is generated by a graph convolution operation from the first-generation graph represented by Eq. 6., and the third-generation graph is generated by a graph convolution operation from the second-generation graph represented by Eq. 7. We subsequently connect them using Eq. 8, where $\tilde{x}_i^{(1)} \parallel \tilde{x}_i^{(2)} \parallel \tilde{x}_i^{(3)}$ represent the extracted feature vectors of the chromosome $i$ in the first, second, and third generation graphs. $\tilde{x}_i^{(merge)}$ refers to the dynamic message passing process among the products represented by chromosome $i$. We connect the bidirectional LSMT layers with a gated attention global pooling layer proposed by Li et al. (2017) to extract high-level information, this model is completed with a dense layer comprising the activation function of sigmoid as a binary classifier in Eq. 9. The gated attention global pooling layer has favorable inductive biases when the problem is graph structured. In this model, by Bishop and Nasrabadi (2006), we use the binary cross entropy as part of the loss function since is a binary classification task. Its definition is in Eq. 21 where $d \in \{0,1\}$ is the true labels of dominant products and $p$ is a probability estimate which can be calculated by $\hat{d}$, the prediction of dominant products in Eq. 20.

$$Accuracy = \frac{TP+TN}{TP+TN+FP+FN}, \text{ Eq. 10}$$

In binary classification cases, the accuracy in Eq. 10 can be a criterion, where TP = True positive; FP = False positive; TN = True negative; FN = False negative. Their definitions are in Eq. 11 to Eq. 14. But our goal is to predict more TP, and the FN might not be very significant to us because identifying a non-dominant product as a dominant product is better than identifying a dominant product as a non-dominant product. Therefore, the importance of TP, TN, FP, and FN is not the same in our case. Their importance order should be $TP > TN$, and $FP > FN$. For the TN and TP values, the more, the better, and for the FN and FP values, the less, the better. In other words, Type I errors are more critical than Type II errors. For our model to be able to trade off their importance, we assign weights to the ratios composed of TP, TN, FP, and FN. Then we add them to the loss function. There are five ratios that we use- true positive rate (TPR), true negative rate (TNR), negative predictive value (NPV), false negative rate (FNR), false positive error rate (FPER), and dispersion rate (DR). The role of DR is to avoid all outputs predicted by the model being of the same kind. Their definitions are provided in Eq. 15 to Eq. 20 and we use them to form a vector $R$ in Eq. 24. We can assign them different weights to control their importance in the total loss. The weights are denoted as $w_{TPR}, w_{TNR}, w_{PPV}, w_{NPV}, and\ w_{FPER}$ and they form a vector $W$ in Eq. 23. Our goal is to minimize the total loss when we train the model. We set the weights as $w_{TPR} = 0.5, w_{TNR} = 0.1, w_{PPV} = 0.1, w_{NPV} = 0.1, w_{FPER} = 0.2, and\ w_{DR} = 0$, adding to 1, to control the importance of the ratios. The numerical setting of these weights depends on the researcher's focus on various aspects. The weights set above are the optimal combination after many experiments. With Eq. 25, by calculating the dot product of $W$ and $R$, we can obtain a scalar $S_w$ which represents an Overall Weighted Score (OWS) that measures the prediction results from the researcher's perspective. In Eq. 26, we define the total loss as the binary cross entropy minus the OWS, because we want to maximize the OWS and minimize the binary cross entropy. Our objective is to learn the optimum parameters $\theta^*$ in the FCP-GNN model, which can minimize the total loss in Eq. 27.

$$TP(d_i, \hat{d}_i) = \sum_{i=0}^{n-1}(d_i = 1)\ and\ (\hat{d}_i = 1),\ \text{Eq. 11}$$

$$FP(d_i, \hat{d}_i) = \sum_{i=0}^{n-1}(d_i = 0)\ and\ (\hat{d}_i = 1),\ \text{Eq. 12}$$

$$TN(d_i, \hat{d}_i) = \sum_{i=0}^{n-1}(d_i = 0) \text{ and } (\hat{d}_i = 0), \text{ Eq. 13}$$

$$FN(d_i, \hat{d}_i) = \sum_{i=0}^{n-1}(d_i = 1) \text{ and } (\hat{d}_i = 0), \text{ Eq. 14}$$

where $i$ refers to the index of chromosome, and $n$ is the total count of chromosomes.

$$TPR(d_i, \hat{d}_i) = \frac{TP}{TP+FN}, \text{ Eq. 15}$$

$$TNR(d_i, \hat{d}_i) = \frac{TN}{TN+FP}, \text{ Eq. 16}$$

$$PPV(d_i, \hat{d}_i) = \frac{TP}{TP+FP}, \text{ Eq. 17}$$

$$NPV(d_i, \hat{d}_i) = \frac{TN}{TN+FN}, \text{ Eq. 18}$$

$$FPER(d_i, \hat{d}_i) = \frac{FP}{FN+FP}, \text{ Eq. 19}$$

$$DR(d_i, \hat{d}_i) = -\frac{(TP-FN+FP-TN)^2}{(FN+FP+TP+TN)^2}, \text{ Eq. 20}$$

$$p_i = \Pr(\hat{d}_i = 1), \text{ Eq. 21}$$

$$\text{Binary cross entropy}: L_{log}(d_i, p_i) = -\frac{1}{n}\sum_{i=0}^{n-1}(d_i \log(p_i) + (1-d_i)\log(1-p_i)), \text{ Eq. 22}$$

$$W = [w_{TPR} \quad w_{TNR} \quad w_{PPV} \quad w_{NPV} \quad w_{FPER} \quad w_{DR}], \text{ Eq. 23}$$

$$R = \begin{bmatrix} TPR \\ TNR \\ PPV \\ NPV \\ FPER \\ DR \end{bmatrix}, \text{ Eq. 24}$$

$$\text{Overall weighted score}: S_w = W \cdot R, \text{ Eq. 25}$$

$$Total\ loss\colon L_{total}(d_i, \hat{d}_i) = L_{log}(d_i, p_i) - S_w, \text{ Eq. 26}$$

$$Objective\colon \theta^* = \underset{\theta}{\operatorname{argmin}} \frac{1}{n} \sum_{i=0}^{n-1} L_{total}(d_i, f_{PFCN}(x_i, \theta)), \text{ Eq. 27}$$

As the dominant product's count is much less than the non-dominant ones, our dataset is imbalanced. We balanced it by dominant and non-dominant product class weights ratio, when the model calculates the loss in the training progress. We set each class weight as the inverse of its class size.

After we constructed the FCPN graph, we detected the labels of dominant products and generate the masks that can split the dataset. Then, we fed this information into the GNN to obtain embedding vectors and generated dynamic graph vectors. We then fed them into a bidirectional LSMT structured DNN to train the model. The entire framework is presented in Fig. 5. [Fig. 5 near here]

## 5. Results

### 5.1. Dominant technologies

The identification of dominant technologies depends on the threshold of the total product count ratio. In this research, we use 50% of the total product count ratio as the threshold of 0.5. However, for the robustness of the dominant technology identification, we also consider 40% and 60 % of the product count ratio as thresholds of 0.4, and 0.6 to identify them. The dataset is split into training, validation, and test datasets in chronological order, which can influence the count of dominant technologies in each dataset by different splitting. When the threshold is 0.4, 0.5, and 0.6, there are 11,126 technologies, out of which 209, 161, and 125, respectively, are dominant. The mean years for a genotype to become dominant are approximately 3.14, 3.4, and 4 years. Therefore, we choose 3 years as our test dataset period and in the GNN, we construct the time series for three generations in the bidirectional LSMT layer. The maximum period for becoming dominant is 14 years and the minimum period is 0. A detailed description is presented in Table 2. Fig. 6 shows the distribution of time taken by a genotype to become dominant by year when the threshold equals 0.5. The y-axis is the time taken and is represented along the y-axis; the x-axis shows the specific year. Fig. 7 indicates the number of years required for genotypes to become dominant at different thresholds. "yeardiff" refers to the years required for the genotypes to become dominant. [Table 2 near here] [Fig. 6 near here] [Fig. 7 near here]

*5.2.Dominant products*

The number of products has increased from 2001 to 2020; hence, we use the ratio of the number of dominant products to the total number of products per year to represent the change in the number of dominant products under different thresholds. In Fig. 8, we can see that the different threshold curves for the first three years have the same ratio. Moreover, their trends are all decreasing and cyclical, with a period of approximately three years. In the same year, the ratio when the threshold is 0.4 is greater than that when the threshold is 0.5, and the ratio when the threshold is 0.5 is greater than that when the threshold is 0.6. Because the data for the final year of 2021 is incomplete, there will be some bias. When the threshold is 0.4, 0.5, and 0.6, there are 3,430 products, out of which 864, 657, and 415 are dominant products, respectively. Table 3 shows the dominant products and genotype counts by year. In 2016 and 2018, the dominant product ratio is very low as the dominant genotype count is small in those years. However, in 2020, there are 83 dominant products resulting from one dominant genotype, which is a CPU frequency of "1.8GHzA55*4". This genotype appears in 2020 and we identify it dominant because many products in 2021 has this genotype. However, in 2021, we only have 3.5 months of data; this may influence the dominant genotype and dominant product identification. [Fig. 8 near here] [Table 3 near here]

In our definition, a dominant product is a result of a dominant genotype. One dominant genotype can generate more than one dominant product, as, in one year, there may be many products having the same dominant genotype. 错误!未找到引用源。 shows that one dominant genotype is contained approximately 7 to 9 dominant products on average in the thresholds from 0.4 to 0.6. The maximum count of the dominant products that contain the same dominant genotype is 48, 83, and 109. Conversely, one dominant product can be generated from more than one dominant genotype. 错误!未找到引用源。 shows the dominant genotype counts in one dominant product. One dominant product contains approximately 2 dominant genotypes on average and a maximum of 11 when the threshold equals 0.4 and 0.5. [Fig. 9 near here] [Fig. 10 near here]

### 5.3. Predictions

We use different thresholds from 0.4 to 0.6 and different years to split the dataset; hence, we have different counts of dominant products and dominant genotypes in the training dataset, validation dataset, and test dataset. The total years are 21. We split the dataset into training, validation, and test datasets in 6 ways: with the years of (4,14,3), (6,12,3), (9,9,3), (12,6,3), (14,4,3), (16,2,3). Table 4 and Table 5 show the number of dominant products and genotypes in different datasets when the threshold is equal to 0.5. [Table 4 near here] [Table 5 near here]

For evaluating the prediction result, we have several criteria, which include Accuracy and TPR calculated as in Eqs. 10 and 15. We also report the TN, TP, FN, and FP, which are defined in Eq. 11 to Eq. 14. As deep learning models are initialized with randomness, to increase the robustness of the model, we train the model and execute the prediction 10 times with different random seeds for the 6 types of datasets splitting. Subsequently, we obtain the average values for each criterion. From the results in Table 6, at the three thresholds of dominant products, the configuration of the training and validation datasets with 9 and 9 years provide the worst results. The configuration of training and validation datasets with 16 and 2 years has the highest value. [Table 6 near here]

We focus on the dominant products at the threshold of 0.5 with the combinations of training and validation dataset years being (16, 2) and (14,4) as our test dataset year is 3, which is between 2 and 4. For robustness, we train and execute the prediction 20 times again with other different random seeds and obtain the average values for the criteria. The results are presented in Table 7. They are comparable with the results presented in Table 5. From the results in Table 7, in both combinations of datasets, the TPR is high, but the accuracy is low. This indicates that there is a tradeoff between FP and FN. As we want fewer FNs, we must accept more FPs. Our model identifies all the TPs without any FPs in most cases for the (16, 2) combination, that is, 16 training years and 2 validation years. We show two examples of results predicted by our model in Fig. 11. [Table 7 near here] [Fig. 11 near here] They both have FN products. Table 8 shows the details of the FN product results in 16 training years and 2 validation years, as shown in the image on the left of Fig 11. In Table 8, the "Genotype" column shows the genotype index that causes the product to be dominant. The genotype "310" is the technology "1.8GHz A55*4" in the "CPU frequency" class. The 19 products are dominant only because of this genotype. Although there are other dominant products that also contain this genotype, our model only identifies these 19 products as non-dominant products, which are FNs. The "Year" column shows the year when the genotype was created. All the dominant genotypes in

TP products are ["China Unicom 5G (NRTDD)," "Mobile 5G (NRTDD)," "1.8GHzA55*4," "10," "ax," "6," "802.11a," "120Hz," "Dual mode 5G"] from the years 2019 to 2021. These genotypes are the whole dominant genotypes in the test dataset when the threshold equals 0.5. Therefore, although FN products exist, their dominant genotypes are already contained in the TP products' dominant genotype set. TP products' dominant genotype set already contains all the dominant genotypes. We check all the prediction results in the 20 runs. They show the same result that all the FN's dominant genotypes are included in the TP's dominant genotype set, which indicates that our model does not miss any dominant genotypes. The time period of our test dataset is until 2021.3; hence, if we know the labels after 2021.3, there may be some current FPs that can become the dominant products in the future. [Table 8 near here]

For comparison purposes, we also use two conventional machine learning methods, the logistic regression proposed by Yu et al. (2011) and the random forest classifier proposed by Breiman (2001), to predict the dominant products in the case of the threshold being 0.5, training years being 18, and test years being 3. As the dataset is imbalanced, we use the SMOTE method proposed by Chawla et al. (2002) to oversample the data and split the dataset with training data from 2001 to 2018, and test data from 2019 to 2021. With original chromosome information, both logistic regression and random forest classifiers show a high accuracy of 0.774% with default parameters in the scikit-learn library; however, they predict all the products as non-dominant. These two methods have no predictive capability in our case, even though we have balanced the data. The results are shown in Table 9. [Table 9 near here]

**6. Discussion and conclusion**

In this study, we defined the dominant product, and our objective was to use a technological phylogenetic network to predict future dominant products. As Box (1976) aptly stated, 'All models are wrong, but some are useful.' Our model may not be the best, but it certainly has its utility. We built the FCPN with Chinese mobile phones from 2001 to 2021 and predicted the dominant products in the framework of FCP-GNN. The FCPN graph data could be used for predicting dominant products with the GNN. It is an evolutionary structure in the GNN that focuses on node-level prediction. Within the dominant genotype thresholds of 0.4,0.5, 0.6, and many combinations of different years for splitting the training dataset, validation dataset, and test dataset, our model shows good performance for predicting the dominant product. Even though there are false negatives in the results, their dominant genotypes are all in the True Positive dominant product's genotypes. This shows that FCPN with GNN can predict the future dominant design, which in turn can represent the trajectory of technology. Our contribution in this study is that we used FCP-GNN to predict dominant products, which means that we can predict product technology evolution in terms of dominant technologies. Graph learning shows excellent predictive ability, even in the case of high uncertainty of technological development. The technological phylogenetic network does not only help in visualization, but it can also be used as the FCPN, which is a high-quality structural input for GNN. We are the first ones to construct the FCPN for dominant product prediction.

In addition, we arrived at the following conclusions for this framework. First, our GNN model combines two ChebNet convolution layers and LSTM to analyze the evolutionary network. There may be other GNN structures that also have satisfactory performance, such as the graph convolutional network (GCN). In this study, the graph has only 3430 nodes, and we extract the node feature information from one hot encoding by PCA. This makes the GNN training fast. Our model will also perform better than other models if the graph is big with a large number of links and nodes. Second, the FCPN keeps all the links between adjacent generations in a manner similar to a neural network.

This structure does not add computational complexity. However, it reduces the complexity of calculation, in the construction of the conventional phylogenetic tree, we have to perform the same calculations and then retain the link of the high similarity and remove other links. Our algorithm only keeps all the links and does not append any other operation. Moreover, FCPN does not calculate the connections between the non-adjacent generations as that will increase the complexity of the calculation. However, any genotype can be transferred through full connections to any later generations, which has similar effects to links between non-adjacent generations. Third, the link weight is set as the similarity between two nodes, which leads to the graph containing more information in this structure. Fourth, compared with the conventional phylogenetic tree, the FCPN keeps all the structural information of the phylogenetic tree and does not provide any bias to the links between generations. The conventional phylogenetic tree is only a subset of FCPN with bias as it removes links that may contain valuable information. Fifth, the FCP-GNN has better prediction results than the general CNN, RNN, and DNN; moreover, it trains quickly. With this framework, the genotype information can be decreased considerably for the same result. Last, our dataset is a specific dataset; however, this framework can also work well on other datasets of the evolutionary process. In our approach, we use weights metrics to control the importance of criteria. However, the weights are hyperparameters. We need more trials to obtain optimized hyperparameters. Moreover, we only used 10–20 predictions and averaged the values to remove randomness. In the future, we will perform more predictions. Our model's performance indicates that within the mobile phone industry, during the specified period, a focus solely on technology yielded favorable outcomes. However, this approach does not ensure precise forecasts in other industries or during the 1990s, which was a time of exploration preceding the consolidation phase (early 2000s) in the mobile phone sector. For example,

Nesta and Dibiaggio (2003) demonstrate how the search process underwent significant evolution over time in industries that utilized emerging biotechnologies. Our methodology offers a valuable estimate of the likelihood of various technological combinations. More specifically, technological selections are influenced by the degree of complementarity among technologies within their respective application contexts. As a result, as products evolve, design elements tend to stabilize—for instance, the array of applications and tools integrated into a mobile phone—and technological trials uncover the most efficacious technologies for joint utilization in mobile phone development. The algorithm is designed to assess the probability of technology occurrences, discern patterns of the most effective combinations, and forecast technological alternatives for future products.

For future research, we plan to apply a heterogeneous graph that contains both products and genotypes simultaneously, because our focus is on the dominant genotypes that represent technologies rather than only the products. Moreover, there is still considerable scope for improvement in the accuracy of our model. We will attempt to use other structures of GNN and loss functions to improve it. Moreover, if the data are combined with market sales data or survey data, the effect may be better. Finally, we plan to increase the generalizability of the model so that it can work in the condition of OOD (out of distribution) rather than only i.i.d., which implies the model should work well even with considerable differences between the training datasets and test datasets.

Acknowledgment

This research was supported by the BK21 FOUR (Fostering Outstanding Universities for Research) funded by the Ministry of Education (MOE, Korea) and National Research Foundation of Korea (NRF) grant funded by the Korea government (MSIT) under Grant 2022R1A2C1091917.

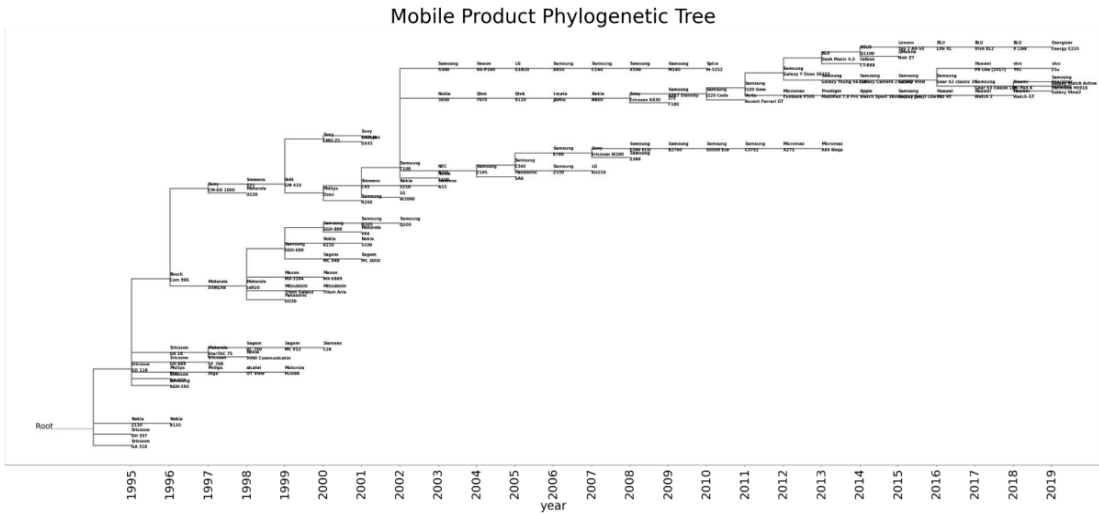

Fig. 1 Example of a conventional product phylogenetic tree

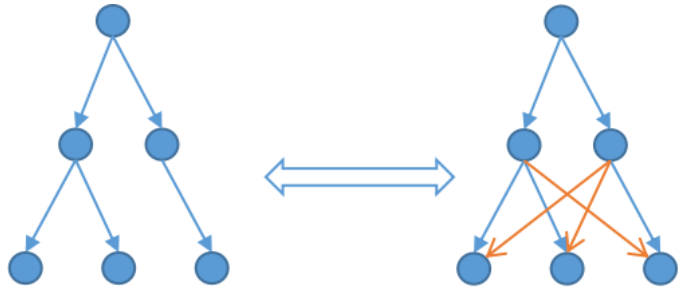

Fig. 2 Example comparison of a conventional phylogenetic tree (left) and a FCPN (right)

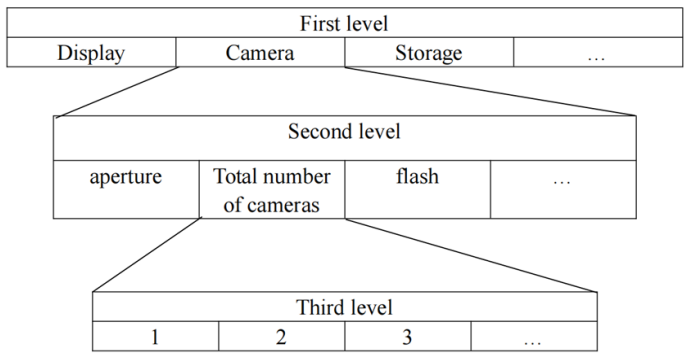

Fig. 3 Example of a 3 levels hierarchical structure of genotype

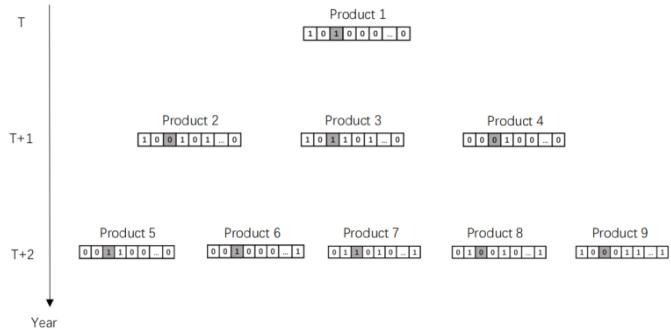

Fig. 4 Example of dominant product identification

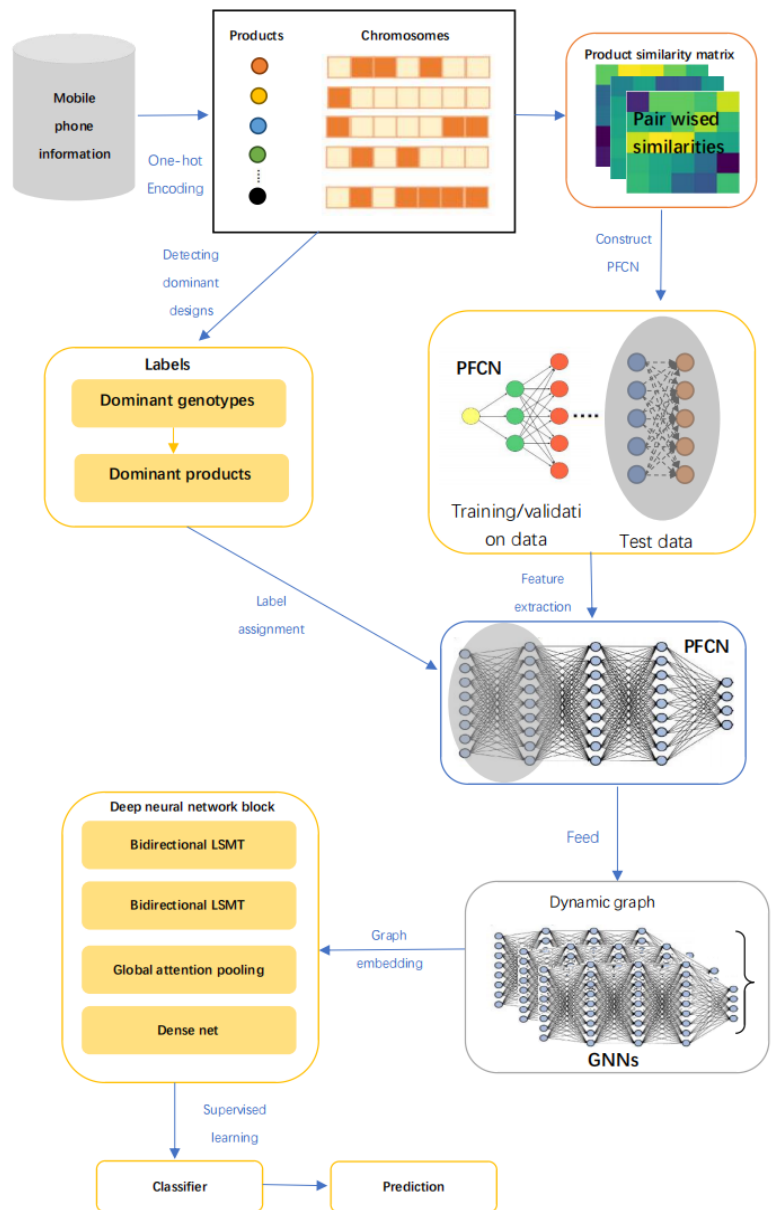

Fig. 5 Framework of the FCP-GNN approach

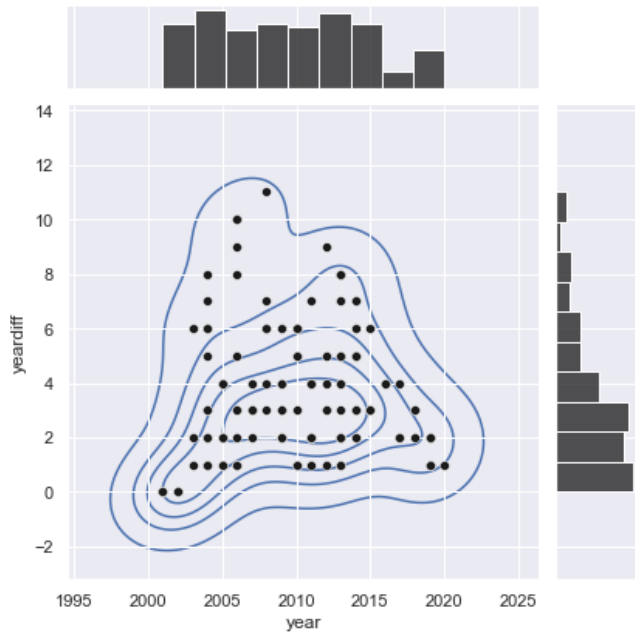

Fig. 6 Distribution of time taken for a genotype to become dominant by year when the threshold equals 0.5

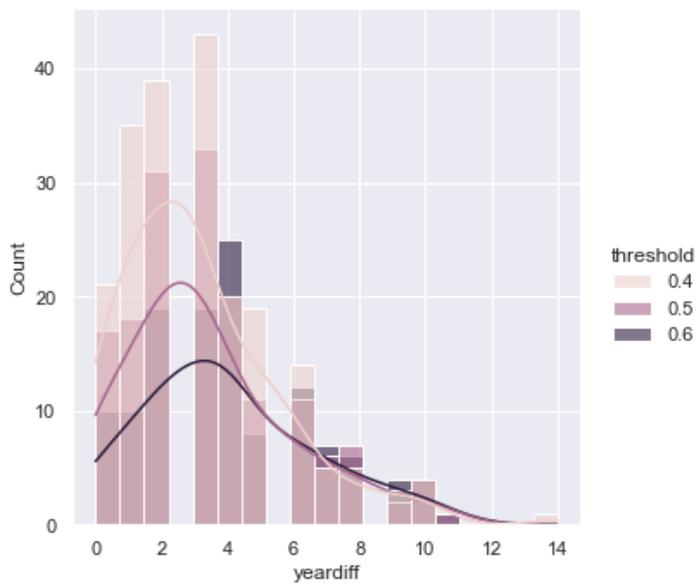

Fig. 7 Number of years required for genotypes to become dominant at different thresholds

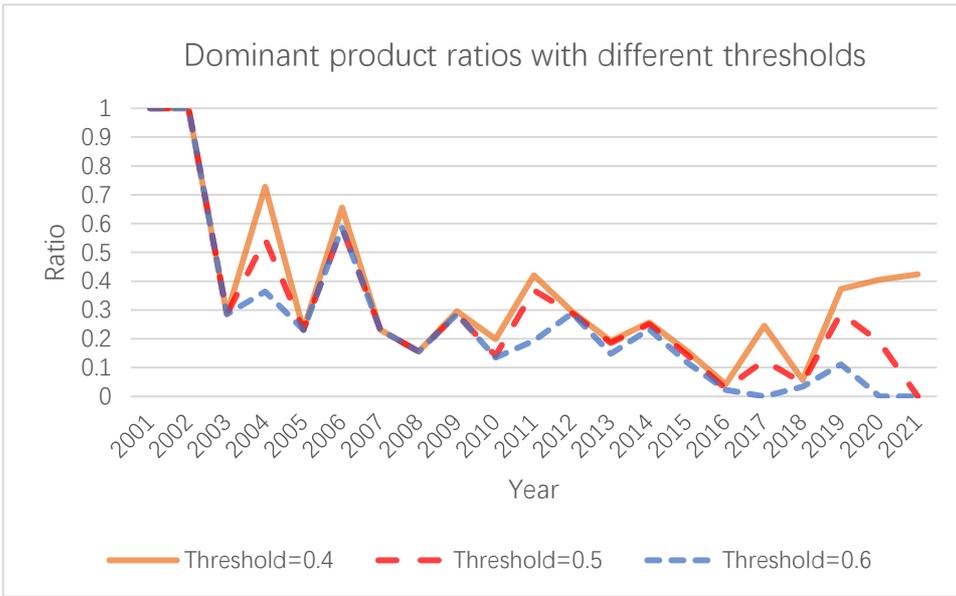

Fig. 8 Dominant product ratios with different thresholds

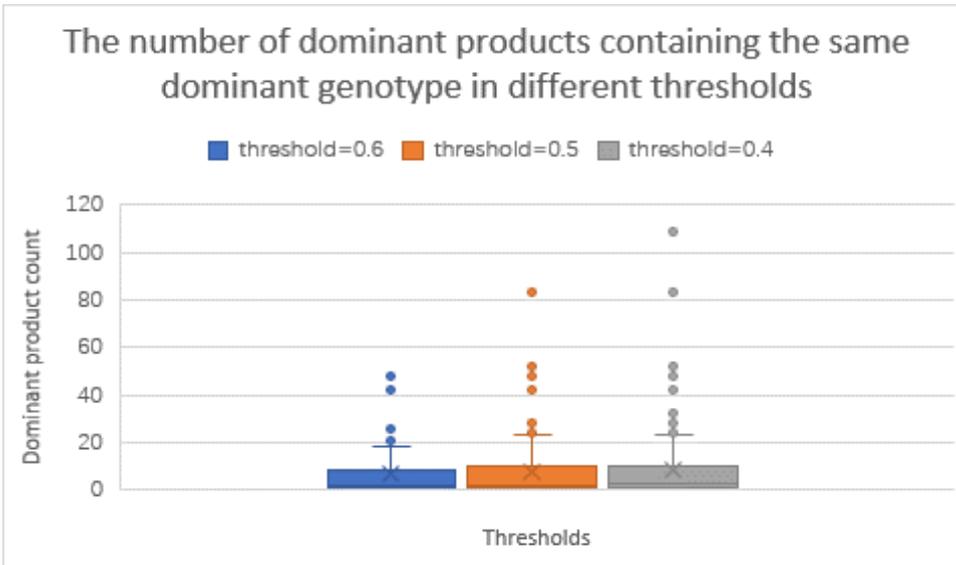

|  | Threshold=0.6 | Threshold=0.5 | Threshold=0.4 |
|---|---|---|---|
| Mean | 8.751 | 7.932 | 6.776 |
| Max | 109 | 83 | 48 |

Fig. 9 Number of dominant products containing the same dominant genotype at different thresholds

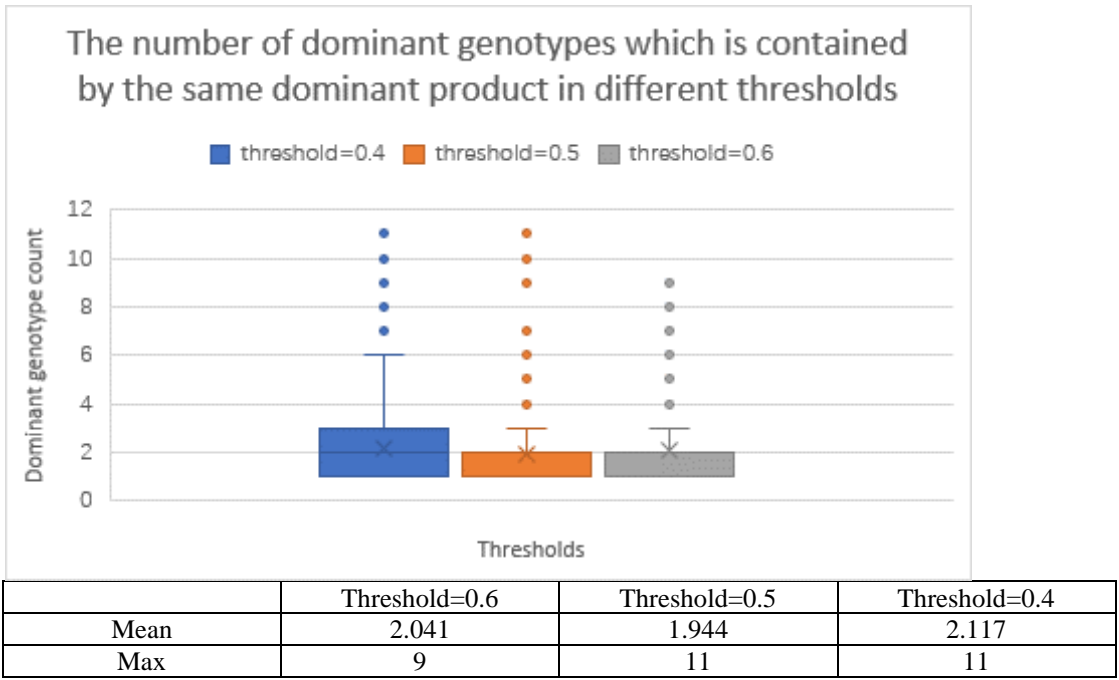

|  | Threshold=0.6 | Threshold=0.5 | Threshold=0.4 |
|---|---|---|---|
| Mean | 2.041 | 1.944 | 2.117 |
| Max | 9 | 11 | 11 |

Fig. 10 The number of dominant genotypes which is contained by the same dominant product in different thresholds

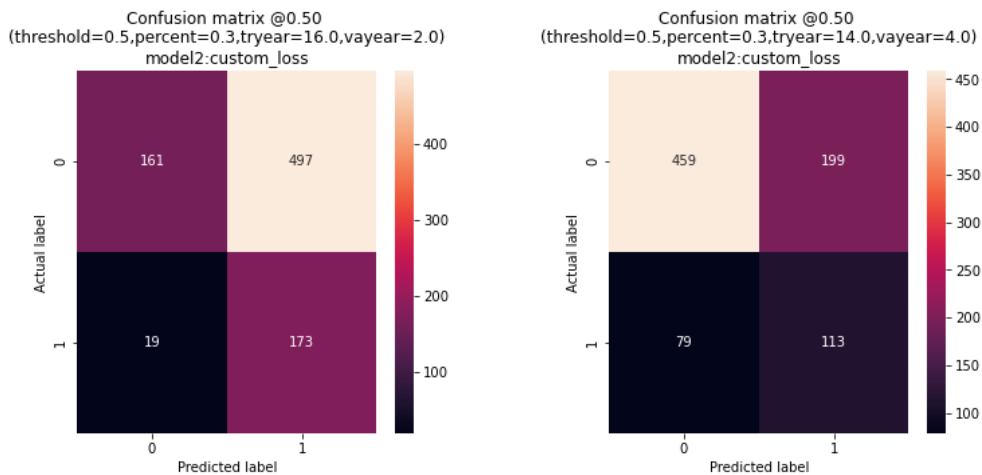

Fig. 11 Two prediction examples in the 20 runs of the predictions when the dominant threshold is 0.5, training years and validation years combinations are (16,2) on the left and (14,4) on the right

| Notations | Descriptions |
| --- | --- |
| $\mathbb{R}^m$ | $m$-dimensional Euclidean space |
| $\alpha, a, A$ | Scalar, vector, and matrix |
| $c_w \star x$ | Convolution function of $c_w$ and $x$ |
| $I_N$ | Identity matrix of dimension N |
| $d_i$ | Dummy variable representing the dominant state of the product $i$. If product $i$ is a dominant product, the value is 1. Otherwise, the value is 0. |
| $g_{n,i}$ | Product $i's\ nth$ genotype: If the product has this genotype, it is 1. If not, it is 0. |
| $x$ | A product's chromosome |
| $\hat{x}$ | Extracted information of a product's chromosome |
| $Cheb()$ | Chebyshev convolutional layer function |
| $\parallel$ | Vector concatenation |
| $\cdot$ | Vector dot product |
| $BiLSTM()$ | Function of bidirectional long-short term memory. |
| $Gap()$ | Gated attention global pooling layer function |
| $PCA()$ | Function of principal components analysis |
| $f_{PFCN}()$ | FCP-GNN model function |
| $\theta$ | Vector of learnable parameters in the model |

Table 1 Notations used in this paper

|  | threshold=0.4 | threshold=0.5 | threshold=0.6 |
|---|---|---|---|
| Dominant genotype count | 209 | 161 | 125 |
| Mean (years) | 3.14 | 3.4 | 4 |
| Std (years) | 2.42 | 2.5 | 2.6 |
| Min (years) | 0 | 0 | 0 |
| 25% (years) | 1 | 2 | 2 |
| 50% (years) | 3 | 3 | 4 |
| 75% (years) | 4 | 5 | 6 |
| Max (years) | 14 | 11 | 11 |

Table 2 Description of time taken for a genotype to become dominant at different thresholds.

| year | Dominant product count | Total product count | Dominant product ratio | Dominant genotype count |
|---|---|---|---|---|
| 2001 | 1 | 1 | 1 | 3 |
| 2002 | 2 | 2 | 1 | 14 |
| 2003 | 2 | 7 | 0.286 | 3 |
| 2004 | 6 | 11 | 0.545 | 12 |
| 2005 | 3 | 13 | 0.231 | 12 |
| 2006 | 17 | 29 | 0.586 | 13 |
| 2007 | 12 | 52 | 0.231 | 5 |
| 2008 | 7 | 45 | 0.156 | 6 |
| 2009 | 28 | 98 | 0.286 | 14 |
| 2010 | 22 | 156 | 0.141 | 8 |
| 2011 | 63 | 171 | 0.368 | 11 |
| 2012 | 56 | 194 | 0.289 | 6 |
| 2013 | 55 | 297 | 0.185 | 17 |
| 2014 | 89 | 355 | 0.251 | 15 |
| 2015 | 54 | 376 | 0.144 | 5 |
| 2016 | 7 | 261 | 0.027 | 2 |
| 2017 | 27 | 216 | 0.125 | 3 |
| 2018 | 14 | 296 | 0.047 | 3 |
| 2019 | 109 | 378 | 0.288 | 8 |
| 2020 | 83 | 439 | 0.189 | 1 |
| 2021 | 0 | 33 | 0 | 0 |

Table 3 Dominant product and genotype distributions at 0.5 threshold

| Years (Training, Validation, Test) | Training dominant products | Training Non-dominant products | Validation dominant products | Validation non-dominant products | Test dominant products | Test non-dominant products |
|---|---|---|---|---|---|---|
| (4,14,3) | 11 | 10 | 454 | 2105 | 192 | 658 |
| (6,12,3) | 31 | 32 | 434 | 2083 | 192 | 658 |
| (9,9,3) | 78 | 180 | 387 | 1935 | 192 | 658 |
| (12,6,3) | 219 | 560 | 246 | 1555 | 192 | 658 |
| (14,4,3) | 363 | 1068 | 102 | 1047 | 192 | 658 |
| (16,2,3) | 424 | 1644 | 41 | 471 | 192 | 658 |

Table 4 Count of dominant products in different datasets when the threshold is 0.5

| Years (Training, Validation, Test) | Training dominant genotypes | Training Non-dominant genotypes | Validation dominant genotypes | Validation non-dominant genotypes | Test dominant genotypes | Test non-dominant genotypes |
|---|---|---|---|---|---|---|
| (4,14,3) | 32 | 35 | 120 | 7086 | 9 | 3844 |
| (6,12,3) | 57 | 127 | 95 | 6994 | 9 | 3844 |
| (9,9,3) | 82 | 547 | 70 | 6574 | 9 | 3844 |
| (12,6,3) | 107 | 1489 | 45 | 5632 | 9 | 3844 |
| (14,4,3) | 139 | 2712 | 13 | 4409 | 9 | 3844 |
| (16,2,3) | 146 | 4750 | 6 | 2371 | 9 | 3844 |

Table 5 Count of dominant genotypes in different datasets when the threshold is 0.5

| Dominant product threshold | Training years | Validation years | TN | TP | FP | FN | Accuracy | TPR |
|---|---|---|---|---|---|---|---|---|
| 0.4 | 14 | 4 | 239.9 | 243.7 | 277.1 | 89.3 | 0.569 | 0.732 |
| 0.4 | 16 | 2 | 82.8 | 308 | 434.2 | 25 | 0.460 | 0.925 |
| 0.4 | 12 | 6 | 288.5 | 219 | 228.5 | 114 | 0.597 | 0.658 |
| 0.4 | 9 | 9 | 457.5 | 69.2 | 59.5 | 263.8 | 0.620 | 0.208 |
| 0.4 | 6 | 12 | 307.8 | 207.2 | 209.2 | 125.8 | 0.606 | 0.622 |
| 0.4 | 4 | 14 | 288.7 | 219.1 | 228.3 | 113.9 | 0.597 | 0.658 |
| 0.5 | 14 | 4 | 379.4 | 128.4 | 278.6 | 63.6 | 0.597 | 0.669 |
| 0.5 | 16 | 2 | 158 | 167.6 | 500 | 24.4 | 0.383 | 0.873 |
| 0.5 | 12 | 6 | 353.4 | 134 | 304.6 | 58 | 0.573 | 0.698 |
| 0.5 | 9 | 9 | 622.9 | 4.6 | 35.1 | 187.4 | 0.738 | 0.024 |
| 0.5 | 6 | 12 | 397.4 | 122.4 | 260.6 | 69.6 | 0.612 | 0.638 |
| 0.5 | 4 | 14 | 424.8 | 116.7 | 233.2 | 75.3 | 0.637 | 0.608 |
| 0.6 | 14 | 4 | 614 | 35.6 | 194 | 6.4 | 0.764 | 0.848 |
| 0.6 | 16 | 2 | 377.9 | 41.2 | 430.1 | 0.8 | 0.493 | 0.981 |

| | | | | | | | |
|---|---|---|---|---|---|---|---|
| 0.6 | 12 | 6 | 555.9 | 39.2 | 252.1 | 2.8 | 0.700 | 0.933 |
| 0.6 | 9 | 9 | 764.8 | 3.7 | 43.2 | 38.3 | 0.904 | 0.088 |
| 0.6 | 6 | 12 | 561.3 | 36.9 | 246.7 | 5.1 | 0.704 | 0.879 |
| 0.6 | 4 | 14 | 546.8 | 39 | 261.2 | 3 | 0.689 | 0.929 |

Table 6 Average values for each criterion of different types of dataset splitting

| Dominant product threshold | Training years | Validation years | TN | TP | FP | FN | Accuracy | TPR |
|---|---|---|---|---|---|---|---|---|
| 0.5 | 16 | 2 | 121.3 | 178.05 | 536.7 | 13.95 | 0.352 | 0.927 |
| 0.5 | 14 | 4 | 347.4 | 134.3 | 310.6 | 57.7 | 0.567 | 0.699 |

Table 6 Average results of predictions executed 20 times when the threshold is 0.5, with different combinations of training years and validation years

| | FN product | Genotype | Year |
|---|---|---|---|
| 0 | Kreta v11v (12gb / 512gb / all Netcom / 5G / alligator) | 310 | 2020 |
| 1 | ASUS zenfone 7 (6GB / 128GB / all Netcom / 5G version) | 310 | 2020 |
| 2 | ASUS zenfone 7 (8GB / 128GB / all Netcom / 5G version) | 310 | 2020 |
| 3 | Lenovo Savior Gaming Phone Pro (12GB/128GB/Full Netcom/5G Version) | 310 | 2020 |
| 4 | Lenovo Savior Gaming Phone Pro (12GB/256GB/Full Netcom/5G Version) | 310 | 2020 |
| 5 | Lenovo Savior Gaming Phone Pro (12GB/512GB/Full Netcom/5G Version) | 310 | 2020 |
| 6 | Lenovo Savior Gaming Phone Pro (8GB/128GB/Full Netcom/5G Version) | 310 | 2020 |
| 7 | Meizu 17 (8GB/128GB/full Netcom/5G version) | 310 | 2020 |
| 8 | Meizu 17 (8GB/256GB/full Netcom/5G version) | 310 | 2020 |
| 9 | Meizu 17 (8GB/256GB/Full Netcom/5G Version/Carrier Limited Edition) | 310 | 2020 |
| 10 | Samsung Galaxy Note 20 Ultra (12GB/256GB/Full Netcom/5G version) | 310 | 2020 |
| 11 | Samsung Galaxy Note 20 Ultra (12GB/512GB/Full Netcom/5G version) | 310 | 2020 |
| 12 | Samsung Galaxy S20 FE (8GB/128GB/Full Netcom/5G version) | 310 | 2020 |
| 13 | One plus 8 (12GB/256GB/full Netcom/5G version) | 310 | 2020 |
| 14 | One plus 8 (8GB/128GB/full Netcom/5G version) | 310 | 2020 |
| 15 | OnePlus 8T (12GB/256GB/Full Netcom/5G version) | 310 | 2020 |
| 16 | OnePlus 8T (12GB/256GB/Full Netcom/5G Version/Cyberpunk 2077 Limited Edition) | 310 | 2020 |
| 17 | One plus 8T (8GB/128GB/full Netcom/5G version) | 310 | 2020 |
| 18 | James V11V (12GB/512GB/full Netcom/5G version/crocodile version) | 310 | 2020 |

Table 7 The detail of the 19 FN products in 错误!未找到引用源。

| | Logistic | Random forest |
|---|---|---|
| TN | 658 | 658 |
| FP | 0 | 0 |
| FN | 192 | 192 |
| TP | 0 | 0 |

Table 8 Prediction results of logistic regression and random forest classifier

**Function**: constructPTPN()

**Input:** All the products with their technological attributes by years t

**Output:** A phylogenetic tree of products

Transforming product information into chromosomes.

Set threshold $\theta$

Build graph $G = (V, E)$

For t in $[T_1, T_2, \ldots, T_n]$

    For $product_{i,t}$ in $[product_{1,t}, product_{2,t}, \ldots, product_{K,t}]$

      $V_t = list()$

        If $product_{i,t}$ is not in $V_t$

          Add node $product_{i,t}$ to $V_t$

          Add node $product_{i,t}$ to $V$

For t in $[T_1, T_2, \ldots, T_n]$

  If $t + 1 \leq T_n$

    For $product_{i,t}$ in $V_t$

     For $product_{j,t+1}$ in $V_{t+1}$

        Calculate $e_{i,j} = Similarity(product_{i,t}, product_{j,t+1})$,

        # Similarity is calculated by product's chromosome

        $E_i = list()$

        If ($e_{i,j}$ not in $E_i$) and ($e_{i,j} \geq \theta$)

      $z = \underset{i}{\operatorname{argmax}}\, e_{i,j}$

        Add link $e_{z,j}$ to $E$

**Return** $G$ (conventional product phylogenetic tree)

---

**Function:** constructTTPN()

**Input:** All the products with their technological attributes by years t

**Output:** A phylogenetic tree of the taxon

Transforming product information into chromosomes.

Set thresholds $\theta, \delta$

For year t in $[T_1, T_2, \ldots, T_n]$

      Do community detection with $\delta$ to all products in year $t$,

    #We cluster product by years to get $[taxon_{1,t}, \ldots, taxon_{n,t}]$

    For $taxon_{i,t}$ in $[taxon_{1,t}, \ldots, taxon_{n,t}]$

      $m = 0$

      For $product_{j,t}$ in $taxon_{i,t}$

        $taxonChro_{i,t}\mathrel{+}= product_{j,t}\text{'s chromosome}$

        $m \mathrel{+}= 1$

      $taxonChro_{i,t} = taxonChro_{i,t}/m$

      '''

      $taxonChro_{i,t}$ is $taxon_{i,t}$'s chromosome which is the average chromosome of all the product's chromosome in the $taxon_{i,t}$

      '''

$TG = constructPTPN(taxonChro)$

# Treat taxon as product and construct the phylogenetic tree

**Return** $TG$ (conventional taxon phylogenetic tree)

---

*The chromosome is an "n*1" vector. Each column represents a technological attribute (genotype). In each column, a binary number (1 or 0) exists indicating whether this product has this genotype.*

*Community detection can be implemented by using the Louvain method (Blondel et al.,2008 ).*

*Threshold $\delta$ is a parameter in the Louvain method to choose the cluster's count, whereas threshold $\theta$ is a filter to choose the links of high-similarity nodes.*

Algorithm 1 Construct the conventional product and taxon phylogenetic trees

**Function:** constructFCPN()

**Input:** All products with their chromosomes grouped by years t

**Output:** A graph of FCPN

Build graph $G = (V, E)$

For t in $[T_1, T_2, \ldots, T_n]$

    For $product_{i,t}$ in $[product_{1,t}, product_{2,t}, \ldots, product_{K,t}]$

      $V_t = list()$

        If $product_{i,t}$ is not in $V_t$

          Add node $product_{i,t}$ to $V_t$

          Add node $product_{i,t}$ to $V$

For t in $[T_1, T_2, \ldots, T_n]$

  If $t + 1 \leq T_n$

    For $product_{i,t}$ in $V_t$

      For $product_{j,t+1}$ in $V_{t+1}$

        Calculate $e_{i,j} = Similarity(product_{i,t}, product_{j,t+1})$,

        # Similarity is calculated by product's chromosome

        If $(e_{i,j}$ not in $E)$

        Add link $e_{i,j}$ to $E$

**Return** $G$

Algorithm 2 Construct the FCPN